\documentclass[12pt]{article}
\usepackage{pmat}
\usepackage{amsmath,amsthm,amsfonts,amssymb}
\usepackage[all]{xy}  
\usepackage{CJK}
\usepackage[whole]{bxcjkjatype} 
\usepackage[unicode]{hyperref} 

\begin{document}
\begin{CJK}{UTF8}{min}

\title{%
Classical Integrable Systems and Gauge Theory\footnote{
この解説は日本の一般向け雑誌の2017年11月号に掲載された以下の記事に基づく：　
浜中 真志, ``可積分系とゲージ場の理論''，
「数理科学」Vol.55-11 (No.653), 49-55.}
}%
\author{
Masashi Hamanaka (Nagoya University)
}

\date{}
 
\baselineskip 6.5mm

\maketitle

\section{What is integrability?}

可積分系とは何だろう？
昔からしばしば取り上げられる問いだが，その答えは
人それぞれである\cite{HSW,Zakharov}.
ただ, これらに共通した性質として
背後に潜む大きな対称性がある.
有限自由度の
Hamilton
系では
Liouville-Arnold
の定理があり，
系の自由度と同じ数の(独立な)保存量が存在すれば
ある種の可積分性が保証される．
これらは通常Noetherの定理を通じて
ラグランジアンの隠れた対称性の
帰結として理解される．
KP方程式，KdV方程式など場の理論の可積分系を統一的に取り扱う
ソリトンの佐藤理論においては，
解空間の持つ無限次元の対称性が可積分性の起源であり，
無限次元代数の言葉で見事に記述される. 
これは運動方程式の対称性であってラグランジアンの対称性ではない. 
Lagrange形式での記述は可能だろうか？
Noetherの定理の帰結として無限個の保存量が導出できるのであろうか？
アノマリーという現象はあるのだろうか？
もっと別の視点から
統一的に理解する枠組みはないであろうか？

本稿では, 
ゲージ場の理論と(古典)可積分系との不思議なつながりについて紹介し,
ツイスター理論の枠組みから可積分性の起源を考察する.
紙数の制約のため原論文への引用をかなり省略した.
(興味ある方は参考文献 
の巻末論文リストなどをご参照ください.)

\section{Ward's inspiration}

1985年にR.Ward氏は, 
可積分系とゲージ理論を結ぶ大変興味深い予想をつぶやいた：

\vspace{2mm}
\noindent
「可積分あるいは可解とみなされる方程式の多く
(ひょっとするとすべて？)は，
反自己双対なゲージ場の方程式(あるいはその一般化)から
リダクションによって得られるかもしれない.」\cite{Ward}

\vspace{2mm}

\noindent
ここで言う「反自己双対なゲージ場の方程式」は
4次元の反自己双対Yang-Mills方程式のことである.
この方程式に対してはツイスター理論での取扱いが
大きな成功を収めており, 
ツイスター理論の可積分系への応用という
新しい潮流が湧き起こった. 
その後の精力的な研究により, 
主要な可積分系のほとんどが実際に
リダクションから得られることが分かり,
1996年出版のMason＆Woodhouseのモノグラフ\cite{MaWo}には, 
ツイスター理論の取扱いも含めて非常に多くの具体例が
ぎっしりと体系的にまとめられている.

\section{Anti-self-dual Yang-Mills equation}

まず主役の反自己双対Yang-Mills方程式を可積分系の枠組みから定義する.
4次元時空の座標を複素化して$z,w,\widetilde{z},\widetilde{w}$と書く. 
計量を$ds^2=dzd\widetilde{z}-dwd\widetilde{w}$で与えれば, 
実スライス：$\widetilde{z}=\overline{z},
\widetilde{w}=-\overline{w}$の場合がEuclid空間, 
$\widetilde{z}=\overline{z},\widetilde{w}=\overline{w}$
あるいは$\widetilde{z},z,\widetilde{w},w\in \mathbb{R}$
の場合が不定値計量$(\!+\!+\!--\!)$の空間に対応する. 
ゲージ群を$G=GL(N,\mathbb{C})$とし
共変微分を, $D_w=\partial_w +A_w$のように表す. 
($A_w(x)$はゲージ場で, $N\times N$行列値関数.
引き数の$x$は4次元座標を表す.)

ここで以下の線形系を考える. 
\begin{eqnarray}
L \varphi&:=&(D_w-\zeta D_{\widetilde{z}}) \varphi=
 0,\nonumber\\
M \varphi&:=&(D_z-\zeta D_{\widetilde{w}}) \varphi=
 0.
\label{lin_asdym}
\end{eqnarray}
$\zeta$はスペクトル・パラメーターと呼ばれる複素数である. 
この方程式系は自由度の点で過剰であるが, 両立条件$[L,M]=0$ を課すことで
解$\varphi(x;\zeta)$が$N$個定まる. 
(このとき$\varphi$は$N\times N$とみなされる.)
この条件式の各$\zeta$の係数から
ゲージ場についての次の方程式が得られる
(ここで$F_{\mu\nu}:=\partial_\mu A_\nu-\partial_\nu A_\mu
+[A_\mu,A_\nu]$)：
\begin{eqnarray}
F_{wz}=
0,~~
 F_{\widetilde{w}\widetilde{z}}
=0,~~
F_{z\widetilde{z}}=F_{w\widetilde{w}}.
\label{asdym}
\end{eqnarray}
これが{\bf 反自己双対Yang-Mills方程式}である.
実スライスを取れば, Hodge作用素$*$に関する
反自己双対方程式$*F_{\mu\nu}=-F_{\mu\nu}$と一致することが分かる.
線形系(\ref{lin_asdym})に対するゲージ変換は以下のようになる：
\begin{eqnarray*}
 L\mapsto g^{-1} L g,~
 M\mapsto g^{-1} M g,~~~
 \varphi\mapsto g^{-1} \varphi,g(x)\!\in\! G.
\end{eqnarray*}
式(\ref{lin_asdym})は広い意味での{\bf Lax形式}と
みなすことができ, Ward予想の一つの根拠となっている.
自己双対Yang-Mills方程式の解はいつも
Yang-Mills理論の運動方程式(Yang-Mills方程式)を満たす.


ここで反自己双対Yang-Mills方程式の対称性についてコメントする.
まず時空の対称性として，
「4次元回転」の対称性と4次元並進の対称性の他に
スケール変換と特殊共形変換の対称性が加わる．
この実15次元の対称性は{\bf 共形対称性}と呼ばれる．
内部空間の対称性としては{\bf ゲージ対称性}がある. 
こちらは空間の各点ごとに決まる局所対称性であり，
巨大な無限次元の対称性である.

\section{Reduction to lower-dimension}

低次元可積分系へのリダクションの例をいくつか紹介する\cite{MaWo}. 
ここで言うリダクションとは, 
次元還元だけでなく,
ゲージ場に対する非自明な拘束条件を含む.
なお，Euclid計量でリダクションすると
時空座標が複素のまま残って困ることがあるので，
通常は不定値計量を採用する\footnote{
なお文献\cite{KMN}は不定値計量に特化した4次元幾何学の貴重な本である.
}.

\subsection{KdV, modified KdV, Non-Linear Schr\"odinger equations}

以下の3つの例では,
ゲージ群は$GL(2,\mathbb{C})$ (あるいはその部分群)で考える.
また次の方向に関する並進不変性を課し次元還元を行う：
$X=\partial_w-\partial_{\widetilde{w}}, Y=\partial_{\widetilde{z}}$.
これによりすべての場は, $(t,x)\equiv (z,w+\widetilde{w})$
だけに依存する. $X,Y$方向の微分はゼロになるので
$\varPhi_X:=A_w-A_{\widetilde{w}}, \varPhi_{\widetilde{z}}:=A_{\widetilde{z}}$は
Higgs場とみなされる. 
反自己双対Yang-Mills方程式は次の形になる：
\begin{eqnarray}
\label{asdym2}
&&
\varPhi_{\widetilde{z}}^\prime 
 +[A_{\widetilde{w}},\varPhi_{\widetilde{z}}] =0,\nonumber\\
 &&
 \dot{\varPhi}_{\widetilde{z}}
 +A_{w}^\prime -A_{\widetilde{w}}^\prime
 +[A_z,\varPhi_{\widetilde{z}}]
 -[A_{w},A_{\widetilde{w}}]=0,\nonumber\\
 &&A_z^\prime -\dot{A}_w+[A_w,A_z] =0.
 \end{eqnarray}
($\dot{f}:=\partial f/\partial t$は時間微分, 
$f^\prime:=\partial f/\partial x$は空間微分.)

まず, KdV方程式へのリダクションを行うため，
以下の非自明な条件をゲージ場に課す：
\begin{eqnarray}
 \label{kdv}
A_{\widetilde{w}}\!=\!\left[\begin{array}{cc}0&0\\u/2&0\end{array}\right], ~
A_w\!=\!\left[\!\begin{array}{cc}0&-1\\u&0\end{array}\!\right],
~
\varPhi_{\widetilde{z}}\!=\!\left[\begin{array}{cc}0&0\\1&0\end{array}\right],
A_z\!=\!\frac{1}{4}
\left[\!\begin{array}{cc} u^\prime& -2u \\
 u^{\prime\prime}+2u^2 &-u^\prime
\end{array}\!\right].\nonumber
\end{eqnarray}
反自己双対Yang-Mills方程式(\ref{asdym2})の第1,2式は自明に満たされ,
第3式の$(2,1)$成分からKdV方程式が導出される：
\begin{eqnarray*}
 \dot{u}=\frac{1}{4}u^{\prime\prime\prime}
  +\frac{3}{2}u u^{\prime}.
\end{eqnarray*}

次に，変形KdV 
方程式へのリダクションを議論する. 
$\varPhi_{\widetilde{z}}$の形はKdVと同じものを取り, 
以下の異なる条件を課してみよう：
\begin{eqnarray*}
 \label{mkdv}
&&
 A_{w}\!=\!
  \left[\!\begin{array}{cc}v&-1\\0&-v\end{array}\!\right],
 A_{\widetilde{w}}\!=\!
  -\frac{1}{2} \left[\!\begin{array}{cc}0&0\\v^\prime+v^2&0\end{array}\!
	       \right],~
 A_z=\frac{1}{4}
  \left[\!\begin{array}{cc}v^{\prime\prime}-2v^3&-2v^\prime
   +2v^2\\
	0&-v^{\prime\prime}+2v^3\end{array}\!\right].
\end{eqnarray*}
すると今度は変形KdV方程式が得られる：
\begin{eqnarray*}
 \dot{v}=\displaystyle
\frac{1}{4}v^{\prime\prime\prime}
-\frac{3}{2}v^{\prime} v^2.
\end{eqnarray*}
ここでKdVと変形KdVの関係について
ゲージ理論の立場から考察する. 
以下の形のゲージ変換は
$\varPhi_{\widetilde{z}}$を不変に保つ：
\begin{eqnarray}
 \label{g}
 g= \left[\begin{array}{cc}1&0\\g_{21}&1\end{array}\right], 
~~~
\varPhi_{\widetilde{z}}\mapsto g^{-1}\varPhi_{\widetilde{z}}g.
\end{eqnarray}
このゲージ変換(\ref{g})によって, KdVのリダクション条件(\ref{kdv})を
変形KdVのもの(\ref{mkdv})と一致させるには, 
$g_{21}=-v$かつ
\begin{eqnarray}
 \label{miura}
u=v^\prime-v^2
\end{eqnarray}
とすればよいことが分かる. 
(\ref{miura})は{\bf Miura変換}として知られている
有名な変数変換である\footnote{Miura変換は
広い意味ではB\"acklund変換の一例であるが, 
本稿では変数変換と解釈する.}.
この文脈では, Miura変換はゲージ変換であると
理解される.

最後に非線形Schr\"odinger 
方程式へのリダクションを与える. 
今度は違うHiggs場の値を取る
($\varepsilon:=\pm 1$):
\begin{eqnarray*}
 \label{nls}
\varPhi_{\widetilde{z}}=
\frac{\mathrm{i}}{2}\left[\begin{array}{cc}-1&0\\0&1\end{array}\right],~
A_{w}=-\left[\begin{array}{cc}0&\psi\\
	     \varepsilon
	    \overline{\psi}&0\end{array}\right],~
A_{z}=\mathrm{i}\varepsilon
  \left[\begin{array}{cc}
-\psi\overline{\psi}
&-\varepsilon\psi^\prime\\
\overline{\psi}^\prime&
\overline{\psi}\psi
	\end{array}\right],~A_{\widetilde{w}}=O.
\end{eqnarray*}
これは非線形Schr\"odinger方程式を導く：
\begin{eqnarray*}
 \label{nls}
  \mathrm{i}\dot{\psi}+\psi^{\prime\prime}+2
  \varepsilon
  \psi  \overline{\psi}  \psi=0.
\end{eqnarray*}
($\varepsilon = 1$
のとき引力型，
$\varepsilon= -1$
のとき斥力型である.)
なお, これとゲージ同値な可積分系として 
Heisenbergの強磁性体方程式がある. 


\subsection{Boussinesq equation}

ここではゲージ群を$GL(3,\mathbb{C})$とし，
前例(a)とは異なる方向に次元還元を行う:
$X=\partial_{\widetilde{w}},~Y=\partial_{\widetilde{z}}$.
時間空間座標を
$(t,x)\equiv (z,w)$のように同一視すると, 
反自己双対Yang-Mills方程式は次のようになる：
\begin{eqnarray}
 \label{asdym_N}
 &&[\varPhi_{\widetilde{w}},\varPhi_{\widetilde{z}}]=0,~
 A_z^\prime -\dot{A}_w+[A_w,A_z] =0,\nonumber\\
&& \dot{\varPhi}_{\widetilde{z}}-\varPhi^\prime_{\widetilde{w}}
 +[A_z,\varPhi_{\widetilde{z}}]
 -[A_{w},\varPhi_{\widetilde{w}}]=0.
\end{eqnarray}
以下の条件をゲージ場に課す：
\begin{eqnarray*}
&&\varPhi_{\widetilde{z}}=
  \left[\begin{array}{ccc}0&0&0\\0&0&0\\1&0&0\end{array}\right],~
 \varPhi_{\widetilde{w}}=
  \left[\begin{array}{ccc}0&0&0\\1&0&0\\0&1&0\end{array}\right],~
 A_z\!=\!
  \left[\begin{array}{ccc}a&0&-1\\d&b&0\\f&e&c\end{array}\!\right],~
 A_{w}\!=\!
  \left[\!\begin{array}{ccc}0&-1&0\\0&0&-1\\v&u&0\end{array}\!\right],\nonumber\\
&&a=-\frac{2}{3}u,~b=c=\frac{1}{3}u,~d=-\frac{2}{3}u^\prime+v,~
e=-\frac{1}{3}u^\prime+v,~
f=-\frac{2}{3}u^{\prime\prime}+v^\prime.
\end{eqnarray*}
反自己双対Yang-Mills方程式(\ref{asdym_N})は$u$と$v$についての微分方程式を与える.
$v$を消去すると, Boussinesq 方程式が得られる：
\begin{eqnarray*}
\ddot{u}+\frac{1}{3}u^{\prime\prime\prime\prime}+\frac{2}{3}
 (u^2)^{\prime\prime}=0.
\end{eqnarray*}

\subsection{(Affine) Toda field equation}

最後に(アファイン) 戸田場方程式の導出を議論する.
今度はゲージ群を$GL(N,\mathbb{C})$とし，
以下の方向に次元還元を行う：
$X=\partial_w,~Y=\partial_{\widetilde{w}}$.
反自己双対Yang-Mills方程式は以下の形になる：
\begin{eqnarray*}
&& \partial_z{\varPhi}_w+[A_z,\varPhi_w]=0,~ \partial_{\widetilde{z}}\varPhi_{\widetilde{w}}
  +[A_{\widetilde{z}},\varPhi_{\widetilde{w}}]=0,\nonumber\\
&&\partial_z A_{\widetilde{z}}-\partial_{\widetilde{z}} A_z
+[A_z,A_{\widetilde{z}}]
+[\varPhi_{\widetilde{w}},\varPhi_w]=0.
\label{++}
\end{eqnarray*}
ゲージ場に以下の条件を課す．
\begin{eqnarray*}
 A_z&=& 
 \mbox{diag}~(a_1,\cdots, a_N),~
 A_{\widetilde{z}}=
\mbox{diag}~(-\widetilde{a}_1,\cdots, -\widetilde{a}_N),~\nonumber\\
 \varPhi_w&=&\left[\begin{array}{ccccc}0&\phi_1&&&O\\&0&\phi_2&&\\
                 &&0&\ddots&\\&O&&\ddots&\phi_{N-1}\\\epsilon\phi_N&&&&0
              \end{array}\right],~
 \varPhi_{\widetilde{w}}=\left[\begin{array}{ccccc}0&&&&\epsilon\widetilde{\phi}_N\\
                \widetilde{\phi}_1&0&&O&\\
                 &\widetilde{\phi}_2&0&&\\&&\ddots&\ddots&\\O&&&\widetilde{\phi}_{N-1}&0
              \end{array}\right],~
\end{eqnarray*}
$\epsilon$は 0または1 の定数である.
$\epsilon=0$のとき, 上記方程式を書き下し,
$a_i, \widetilde{a}_i$を消去して$u_i=\log (\phi_i\widetilde{\phi}_i)$
とおけば，
戸田場の方程式が得られる. 
\begin{eqnarray*}
 \partial_z\partial_{\widetilde{z}}u_i+\sum_{j}K_{ij}e^{u_j}=0,
\end{eqnarray*}
ただし$K_{ij}$ は$SU(N)$ のCartan 行列. 
$\epsilon=1$のときはアファイン戸田場の方程式が得られる.


\section{B\"acklund transformation}

さて, ここから反自己双対Yang-Mills方程式の
可積分性を議論しよう. 
{\bf B\"acklund変換} (解を解にうつす変換)について,
まずはツイスター理論を持ち出さずに
低次元でもなじみのある形で紹介する. 

反自己双対Yang-Mills方程式と等価な {\bf Yangの方程式}を考える：
\begin{eqnarray}
\label{yang}
 \partial_z(J^{-1}  \partial_{\widetilde{z}} J)-\partial_w (J^{-1} \partial_{\widetilde{w}} J)=0.
\end{eqnarray}
$J$は{\bf Yangの行列}と呼ばれる$N\times N$行列である.
この方程式の解$J$が与えられると, 
$J=\widetilde{h}^{-1}h $のように2つの
$N\times N$行列$h$と$\widetilde{h}$
に分解することで 反自己双対ゲージ場を
以下のように再現することができる(理由は次節)：
\begin{eqnarray}
\label{a}
A_{z}=-(\partial_z h)h^{-1}, ~  A_{w}=-(\partial_w h)h^{-1},~
A_{\widetilde{z}}=-(\partial_{\widetilde{z}}\widetilde{h})\widetilde{h}^{-1},~
A_{\widetilde{w}}=-(\partial_{\widetilde{w}}\widetilde{h})\widetilde{h}^{-1}.
\end{eqnarray}
ゲージ変換は$h\mapsto g^{-1} h , \widetilde{h}\mapsto 
g^{-1} \widetilde{h}$として表される. したがって$J$はゲージ不変である. 

以後ゲージ群は$GL(2,\mathbb{C})$ で考える.
一般性を失わず $J$を以下のようにパラメトライズすることができる：
\begin{eqnarray}
 \label{J}
 J=\left[\begin{array}{cc} p -r q^{-1} s&-r q^{-1}
   \\ q^{-1}  s &q^{-1}\end{array}
   \right].
\end{eqnarray}
このとき, 以下の2種類の変換はB\"acklund変換であり, 
反自己双対Yang-Mills方程式を不変に保つ\cite{CFGY,MaWo}：
\begin{itemize}
\item $\beta$変換: 
  $ p_{\mbox{\scriptsize{new}}}=q^{-1},~
   q_{\mbox{\scriptsize{new}}}=p^{-1},\\
   \partial_{\widetilde{z}} r_{\mbox{\scriptsize{new}}}
  \!\!=\!q^{-1} \!(\partial_w s) p^{-1}\!,~
  \partial_{\widetilde{w}} r_{\mbox{\scriptsize{new}}}
  \!\!=\!  q^{-1} \!(\partial_z s) p^{-1},~
  \partial_w s_{\mbox{\scriptsize{new}}}
  \!\!=\!p^{-1} \!(\partial_{\widetilde{z}} r) q^{-1}\!,~
  \partial_z s_{\mbox{\scriptsize{new}}}
   \!\!=\! p^{-1}\!(\partial_{\widetilde{w}} r) q^{-1}$.
\item $\gamma_0$変換:
$p^{-1}_{\mbox{\scriptsize{new}}}=q-sp^{-1} r,~
q^{-1}_{\mbox{\scriptsize{new}}}=p-r q^{-1} s,~
r^{-1}_{\mbox{\scriptsize{new}}}=r-p s^{-1} q,~
s^{-1}_{\mbox{\scriptsize{new}}}=s-q r^{-1} p$.
\end{itemize}
これらはともに包合的($\beta\circ \beta=id, \gamma_0\circ \gamma_0=id$)
であるが, それらを組み合わせた$\alpha=\gamma_0\circ \beta$は
非自明な変換となる. したがってある解
から出発して, B\"acklund変換の作用によって
一連の解のシリーズが構成(生成)される：
\[
\xymatrix{
R_0\ar@{->}[r]^\alpha\ar@{<->}[dr]_\beta&R_1\ar@{->}[r]^\alpha\ar@{<->}[dr]_\beta&R_2\ar@{->}[r]^\alpha\ar@{<->}[dr]_\beta&R_3\ar@{->}[r]&\cdots\\
&R'_1\ar@{->}[r]_{\alpha'}\ar@{<->}[u]^{\gamma_0}&R'_2\ar@{->}[r]_{\alpha'}\ar@{<->}[u]^{\gamma_0}&R'_3\ar@{->}[r]\ar@{<->}[u]^{\gamma_0}&\cdots
}
\]

ここで生成された解を簡明に記述するため, 
Quasideterminantというものを導入しよう. 
$n\times n$行列$A=(a_{ij})$に対して, 
その逆行列の存在を仮定する.
このとき, $A$の{\bf Quasideterminant}は, 逆行列の要素の逆数として, 
\begin{eqnarray}
\label{def_q}
\vert A \vert_{ij}:=(A^{-1})_{ji}^{-1}=(-1)^{i+j}\frac{\det A}{\det A^{ij}}
\end{eqnarray}
のように($n^2$種類だけ)定義される. 
2つ目の等号はLaplaceの公式による. 
($A^{ij}$は$A$の$i$行と$j$列を除いた行列である.)
このように行列式そのものではなく,
上記の行列式の比(一般には逆行列の要素)
に着目したものが威力を発揮する. 
(特に最終節で述べるように非可換化の際, 顕著である.)

なお行列$A$の要素$a_{ij}$を明示したいときは
添え字$i,j$の代わりに$A$の$(i,j)$成分をボックスで囲む
表記法が便利である：
\begin{eqnarray*}
 \vert A\vert_{ij}=
  \begin{array}{|ccccc|}
   a_{11}&\cdots &a_{1j} & \cdots& a_{1n}\\
   \vdots & & \vdots & & \vdots\\
   a_{i1}& \cdots & {\fbox{$a_{ij}$}}& \cdots & a_{in}\\
   \vdots & & \vdots & & \vdots\\
   a_{n1}& \cdots & a_{nj}&\cdots & a_{nn}
  \end{array}
\end{eqnarray*}

Quasideterminantはさまざまな恒等式を満たす\cite{GGRW}.
(行列をブロック分解した表記であるが, 
小文字の要素は$1\times 1$とする.)
\begin{itemize}
 \item QuasiJacobi恒等式(行列式の比として分母を払うと
通常の行列式のJacobi恒等式に帰着する)：
\begin{eqnarray*}
\label{nc syl}
\!   \begin{vmatrix}
      A&B&C\\
      D&f&g\\
      E&h&\fbox{$i$}
    \end{vmatrix}
= \!
    \begin{vmatrix}
      A&C\\
      E&\fbox{$i$}
    \end{vmatrix} \!-\!
    \begin{vmatrix}
      A&B\\
      E&\fbox{$h$}
    \end{vmatrix}
    \begin{vmatrix}
      A&B\\
      D&\fbox{$f$}
    \end{vmatrix}^{-1}\!
    \begin{vmatrix}
      A&C\\
      D&\fbox{$g$}
    \end{vmatrix}.
\end{eqnarray*}
\item ホモロジカル関係式(非可換化すれば非自明で有用)：
\begin{eqnarray*}
\label{row hom}
    \begin{vmatrix}
      A&B&C\\
      D&f&g\\
      E&\fbox{$h$}&i
    \end{vmatrix}\!
=  \!\begin{vmatrix}
      A&B&C\\
      D&f&g\\
      E&h&\fbox{$i$}
    \end{vmatrix}
    \begin{vmatrix}
      A&B&C\\
      D&f&g\\
      0&\fbox{0}&1
    \end{vmatrix},~~
\label{col hom}
    \begin{vmatrix}
      A&B&C\\
      D&f&\fbox{$g$}\\
      E&h&i
    \end{vmatrix} \!
=  \!    \begin{vmatrix}
      A&B&0\\
      D&f&\fbox{0}\\
      E&h&1
    \end{vmatrix}
    \begin{vmatrix}
      A&B&C\\
      D&f&g\\
      E&h&\fbox{$i$}
    \end{vmatrix}
\end{eqnarray*}
\end{itemize}

さて本題に戻ろう. $R_0$に属する種子となる解を見つけるため, 
$p=q=r=s=\varDelta_0^{-1}$ ($\varDelta_0(x)$はスカラー関数)とおくと,
反自己双対Yang-Mills方程式は線形微分方程式
(Euclid空間のときは4次元Laplace方程式)
\begin{eqnarray}
 \label{laplace}
(\partial_z\partial_{\widetilde{z}}
-\partial_w\partial_{\widetilde{w}})\varDelta_0=0
\end{eqnarray}
となる. 
この解$\varDelta_0$に
B\"acklund変換を$l$回施して得られる$R_l$に属する解は
以下のようにQuasideterminantで極めて簡明に表される\footnote{
なお $p_l$と $q_l$は実は同じものであるが, 
非可換化すると区別される.}：
\begin{eqnarray}
\label{AWsol}
p_l&=&
\begin{array}{|cccc|}
\fbox{$\varDelta_0$}&\varDelta_{-1} & \cdots & \varDelta_{-l}\\
\varDelta_1 &\varDelta_0&\cdots & \varDelta_{1-l} \\
\vdots &\vdots &\ddots & \vdots\\
\varDelta_{l} &\varDelta_{l-1} &\cdots &\varDelta_0
\end{array}^{-1},~
q_l=
\begin{array}{|cccc|}
\varDelta_0&\varDelta_{-1} & \cdots & \varDelta_{-l}\\
\varDelta_1 &\varDelta_0&\cdots & \varDelta_{1-l} \\
\vdots &\vdots &\ddots & \vdots\\
\varDelta_{l} &\varDelta_{l-1} &\cdots &\fbox{$\varDelta_0$} 
\end{array}^{-1},\nonumber\\
r_l&=&
\begin{array}{|cccc|}
\varDelta_0&\varDelta_{-1} & \cdots & \varDelta_{-l}\\
\varDelta_1 &\varDelta_0&\cdots & \varDelta_{1-l} \\
\vdots &\vdots &\ddots & \vdots\\
\fbox{$\varDelta_{l}$} &\varDelta_{l-1} &\cdots &\varDelta_0 
\end{array}^{-1},~
s_l=
\begin{array}{|cccc|}
\varDelta_0&\varDelta_{-1} & \cdots & \fbox{$\varDelta_{-l}$}\\
\varDelta_1 &\varDelta_0&\cdots & \varDelta_{1-l} \\
\vdots &\vdots &\ddots & \vdots\\
\varDelta_{l} &\varDelta_{l-1} &\cdots &\varDelta_0
\end{array}^{-1}.
\end{eqnarray}
行列の要素として現れたスカラー関数$\varDelta_i(x)$は
以下の関係式により, $\varDelta_0$から逐次求まる：
\begin{eqnarray}
&&\label{chasing}
 \frac{\partial \varDelta_i}{\partial z}
= -\frac{\partial \varDelta_{i+1}}{\partial \widetilde{w}},~~~
 \frac{\partial \varDelta_i}{\partial w}
= -\frac{\partial \varDelta_{i+1}}{\partial \widetilde{z}},\nonumber\\
&&-l\leq i\leq l-1~~~(l\geq 2),
\end{eqnarray}
$R_l^\prime$に属する解も同様の美しい形をしている.

証明は,(Quasi)determinantの恒等式のみを駆使して与えられる.
例えば, 
$\gamma_0$変換$q_{l}^{-1}=
p^\prime_l-r^\prime_l q^{\prime -1}_l  s^\prime_l$は具体的には,
\begin{eqnarray*}
\begin{array}{|cccc|}
\!\varDelta_0\!&\varDelta_{-1}\! & \!\cdots\! & \varDelta_{-l}\!\\
\!\varDelta_1 \!&\varDelta_0\!&\!\cdots\! & \varDelta_{1-l}\! \\
\!\vdots \!&\vdots \!&\!\ddots \!& \vdots\!\\
\!\varDelta_{l}\!&\varDelta_{l-1}\! &\!\cdots\! &\fbox{$\varDelta_0$}\! 
\end{array}
=
\begin{array}{|ccc|}
\!\varDelta_0\!&\!\cdots \!& \!\varDelta_{1-l}\!\\
\!\vdots \!&\!\ddots \!&\! \vdots\!\\
\!\varDelta_{l-1} \!&\!\cdots \!&\!\fbox{$\varDelta_0$} \!
\end{array}
-\begin{array}{|ccc|}
\!\varDelta_1\!&\!\cdots \!& \!\varDelta_{2-l}\!\\
\!\vdots \!&\!\ddots \!& \!\vdots\!\\
\!\fbox{$\varDelta_{l}$}\! &\!\cdots \!&\!\varDelta_1\! 
\end{array}~
\begin{array}{|ccc|}
\!\fbox{$\varDelta_0$}\!&\!\cdots \!&\! \varDelta_{1-l}\!\\
\!\vdots \!&\!\ddots\! &\! \vdots\!\\
\!\varDelta_{l-1} \!&\!\cdots\! &\!\varDelta_0 \!
\end{array}^{-1}
\begin{array}{|ccc|} 
\!\varDelta_{-1}\! &\!\cdots \!&\! \fbox{$\varDelta_{-l}$} \!\\
\!\vdots \!&\!\ddots \!& \!\vdots\!\\
\!\varDelta_{l-2} \!&\!\cdots \!&\!\varDelta_{-1} \!
\end{array}
\nonumber
\end{eqnarray*}
と等価であるが, これはまさにQuasiJacobi恒等式そのものである！
(コーナーの4成分に着目してQuasiJacobi恒等式を適用すればよい.)
$\beta$変換の証明も, (Quasi)determinantの恒等式を駆使することで
示される\cite{GHN}. したがって「B\"acklund変換は
(Quasi)determinant の恒等式そのものである」と
言い表すこともできるが, これは低次元可積分系で
よく知られた事実である. (非可換化しても通用する.)

線形方程式(Laplace方程式) (\ref{laplace})を解くことは難しくない.
例えば, Euclid計量では基本解として
$\varDelta_0=1+\sum_{i=1}^{k}(\lambda_i/(z\widetilde{z}-w\widetilde{w}))$ 
($\lambda_i$は定数)が取れるが，これは
インスタントン解を与える. 
また，次のような解 
$\varDelta_0=1+c\exp(az+b\widetilde{z}+a w+b \widetilde{w})$
($a,b,c$は定数)は「非線形平面波」解\cite{deVega}と呼ばれるソリトン解を与える. 
これらは局所解であり多様な解が含まれる.
D ブレーン解釈も興味深い問題である. 

\section{Penrose-Ward tranformation}

反自己双対Yang-Mills方程式の可積分性について詳しい議論を行うため
線形系(\ref{lin_asdym})について補足をする.
非自明なゲージ場を得るには, $\varphi$は
$\zeta=\infty$で正則であってはならない. 
したがって $\widetilde{\zeta}=1/\zeta$ の座標変換を施し
$\widetilde{\zeta}=0$の付近では別の線形系を考える：
\begin{eqnarray}
&&
\widetilde{\zeta} D_w  \widetilde{\varphi}-D_{\widetilde{z}} \widetilde{\varphi}=0,\nonumber\\
&& \widetilde{\zeta}D_z \widetilde{\varphi}-D_{\widetilde{w}}
 \widetilde{\varphi}= 0.
\label{lin_asdym2}
\end{eqnarray}
両立条件から反自己双対Yang-Mills方程式が出る. 
(このとき$\widetilde{\varphi}$は$N\times N$行列.)

$\varphi$と$\widetilde{\varphi}$をそれぞれ
$\zeta$と$\widetilde{\zeta}$について原点の周りで展開したときの
0次の部分がそれぞれ(前節で$J$を分解したときに定義した) 
$h(x)$と$\widetilde{h}(x)$である. 
線形系(\ref{lin_asdym})において$\zeta=0$とおくと
$D_z h=0, D_w h=0$より, (\ref{a})の第1行目の式が得られ,  
線形系(\ref{lin_asdym2})において$\widetilde{\zeta}=0$とおくと
(\ref{a})の第2行目の式が得られる.

さて, ここからツイスター理論の枠組み
と解構成法であるPenrose-Ward変換を議論する
(詳しくは参考文献\cite{Dunajski,MaWo,Takasaki}参照). 
{\bf Penrose-Ward変換}とは, 
4次元空間上の反自己双対Yang-Mills方程式の解と
実6次元ツイスター空間上の正則ベクトル束との
1対1対応を与える変換のことである. 

時空の座標$(z,\widetilde{z},w,\widetilde{w})$と
ツイスター空間の局所座標$(\lambda,\mu,\zeta)$とは
以下の付帯関係式で結ばれている：
\begin{eqnarray}
\label{incidence}
\lambda=\zeta w+\widetilde{z},~\mu=\zeta z+\widetilde{w}.
\end{eqnarray}
これよりツイスター関数 $f(\lambda,\mu,\zeta)$ は
以下を満たすことが分かる：
\begin{eqnarray*}
&&lf(\lambda,\mu,\zeta):=(\partial_w-\zeta\partial_{\widetilde{z}})f(\lambda,\mu,\zeta)=0,\nonumber\\
&&mf(\lambda,\mu,\zeta):=(\partial_z-\zeta\partial_{\widetilde{w}})f(\lambda,\mu,\zeta)=0.
\label{hol}
\end{eqnarray*}

ツイスター空間は2枚の局所座標で覆うことができ, 
正則ベクトル束は1種類の変換関数$P$から記述される.
この$P$から反自己双対ゲージ場を構成する方法は以下の通り.
まずBirkhoffの分解定理により, 以下を満たす
$N\times N$行列$\varphi, \widetilde{\varphi}$が存在する：
\begin{eqnarray*}
\label{birkhoff}
 P(\zeta w+\widetilde{z},\zeta z+\widetilde{w},\zeta)=\widetilde{\varphi}^{-1}(x;\zeta) \varphi(x;\zeta).
\end{eqnarray*}
この$\varphi, \widetilde{\varphi}$は線形系(\ref{lin_asdym}),
(\ref{lin_asdym2})を満たすことが分かり, 
$\zeta, \widetilde{\zeta}$展開の0次部分から
$h, \widetilde{h}$を取り出すことで, 反自己双対ゲージ場を
(\ref{a})のように構成できる.

\section{Atiyah-Ward ansatz solutions}

再びゲージ群は$GL(2,\mathbb{C})$とし,
具体解を構成しよう. このとき 
変換関数を以下の形に取るのが便利である：
\begin{eqnarray*}
 P_l(x;\zeta)=\left[\begin{array}{cc}0&\zeta^{-l}  
      \\\zeta^{l} &\varDelta(x;\zeta)\end{array}
   \right],~~~l=0,1,2,\dots.
\end{eqnarray*}
これを$l$次の{\bf Atiyah-Ward仮設}と呼ぶ\footnote{通常は
$\zeta^l, \zeta^{-l}$を
非対角ではなく対角に取るが, 後述のB\"acklund変換の議論で
使いやすい形にした. (本質は不変) }.
これに対応するゲージ場の解が5節のB\"acklund変換で生成された
解(\ref{AWsol})と一致することを示す.

ここで, 変換関数がツイスター関数：
$P_l=P_l(\zeta w+\widetilde{z},\zeta z+\widetilde{w},\zeta)$
であることに注意しよう.  $\varDelta(x;\zeta)$を
以下のように$\zeta$についてLaurent展開すると,  
\begin{eqnarray*}
 \varDelta(x;\zeta) = \sum_{i=-\infty}^{\infty}\varDelta_i(x) \zeta^{-i},
\end{eqnarray*}
$(\partial_w-\zeta\partial_{\widetilde{z}})\varDelta=0,~(\partial_z-\zeta
\partial_{\widetilde{w}}) \varDelta=0$の$\zeta$の各べきから
係数$\varDelta_i(x)$に関する関係式(\ref{chasing})が出てくる.

次に分解問題(Riemann-Hilbert問題) $\widetilde{\varphi}P_l= \varphi$
を解こう：
\begin{eqnarray*}
\left[\!\begin{array}{cc} \widetilde{\varphi}_{11}&\widetilde{\varphi}_{12}
            \\ \widetilde{\varphi}_{21}&\widetilde{\varphi}_{22}\end{array}
   \!\right]
\left[\!\begin{array}{cc}0&\zeta^{-l} 
      \\ \zeta^l & \varDelta(x;\zeta)\end{array}
   \!\right]
\!=\!
\left[\!\begin{array}{cc}\varphi_{11}&\varphi_{12}
      \\ \varphi_{21}&\varphi_{22}\end{array}
   \!\right],
\end{eqnarray*}
すなわち, 
\begin{eqnarray}
\label{splitting1}
&& \widetilde{\varphi}_{12}\zeta^l=\varphi_{11},~~~
\widetilde{\varphi}_{22}\zeta^l=\varphi_{21},\\
&&\widetilde{\varphi}_{11} \zeta^{-l}
+ \widetilde{\varphi}_{12}\varDelta 
=\varphi_{12},~
\widetilde{\varphi}_{21} \zeta^{-l} + \widetilde{\varphi}_{22}\varDelta
=\varphi_{22}.\nonumber
\end{eqnarray}
$\varphi$と$\widetilde{\varphi}$をそれぞれ
$\zeta$と${\widetilde{\zeta}}=\zeta^{-1}$で展開し, 
(\ref{splitting1})の第1行目の式に代入すると,
$\varphi_{11}, \varphi_{21},\widetilde{\varphi}_{12} ,\widetilde{\varphi}_{22}$が
有限項で切れることが分かる. 
それを(\ref{splitting1})の第2行目の式に代入すると, 
$\zeta^{0}, ~\zeta^{-1},~ \cdots, ~\zeta^{-l}$の係数から, 
 $h$ と $\widetilde{h}$ の各成分を含む線形関係式が得られる：
\begin{eqnarray*}
&&(h_{11}, *,\cdots\!,*,\widetilde{h}_{12})
D_{l+1}\!=\!(-\widetilde{h}_{11},0,\cdots\!,0,h_{12}),\nonumber\\
&&(h_{21}, *,\cdots\!,*,\widetilde{h}_{22})
D_{l+1}\!=\!(-\widetilde{h}_{21},0,\cdots\!,0,h_{22}),
\nonumber\\
&&{\mbox{ただし}} D_{l+1}:=
\left[
\begin{array}{cccc}
\varDelta_0&\varDelta_{-1} & \cdots & \varDelta_{-l}\\
\varDelta_1 &\varDelta_0&\cdots & \varDelta_{1-l} \\
\vdots &\vdots &\ddots & \vdots\\
\varDelta_{l} &\varDelta_{l-1} &\cdots &\varDelta_0 
\end{array}
\right].
\end{eqnarray*}
こうして自己双対性の非線形問題が線形問題に帰着した．
あとは, $D_{l+1}$の逆行列を右から掛ければ, 
第$1,l+1$列目から$h$ と $\widetilde{h}$だけの4つの関係式が得られる.
ここで$J$の分解を以下のようにゲージ固定すれば
未知変数は4個となって, (\ref{AWsol})の形の解が
得られる\footnote{符号のずれる部分があるが，
B\"acklund変換の再定義で吸収される.}：
\begin{eqnarray*}
\left[\!\begin{array}{cc} p -r q^{-1} s&-r q^{-1}
   \\ q^{-1}  s &q^{-1}\end{array}
   \!\right]
\!=\!\underbrace{\left[\!\begin{array}{cc}1&r
   \\ 0 &q\end{array}
   \!\right]^{-1}}_{\widetilde{h}^{-1}}\!
\underbrace{\left[\!\begin{array}{cc}
	     p&0
\\ s&1\end{array}
   \!\right]}_{h}.
\label{M-W}
\end{eqnarray*}
したがって5節のB\"acklund変換で生成される解は
Atiyah-Ward仮設解と呼ばれるべきものである. 

最後にB\"acklund変換$\beta, \gamma_0$の起源も説明しよう. 
これらは変換関数$P$への以下の作用として理解される\footnote{正確には$B$の
作用は$\varphi,\widetilde{\varphi}$へのある特異ゲージ変換と合わせて定義される\cite{GHN}.}：
\begin{eqnarray*}
 \beta&:& P_{\mbox{\scriptsize{new}}} = B^{-1} P B,~~~
 \gamma_0: P_{\mbox{\scriptsize{new}}} = C_0^{-1} P  C_0,\nonumber\\
&& B=\left[\begin{array}{cc}0&1\\\zeta^{-1}&0\end{array}\right],~~~
 C_0=\left[\begin{array}{cc}0&1\\1&0\end{array}\right].
\end{eqnarray*}
$\beta\circ \beta=id, \gamma_0\circ \gamma_0=id$は明らかであり,
これらの合成が
$l$次の Atiyah-Ward 仮設を $(l+1)$次の Atiyah-Ward 仮設に
うつすこともすぐに分かる：
$A:=BC_0$として$\alpha: A^{-1}P_lA=P_{l+1}$.
なお, $\gamma_0$変換は
$C_0$の代わりに任意の定数行列$C$を用いた形($\gamma$変換)に
容易に拡張することができる: 
$ \gamma: P_{\mbox {\scriptsize{new}}} = C^{-1} P  C$. 
$\beta$変換と$\gamma$変換全体はループ群の作用$LGL(2)$を与える．
これがすべての解を生成するかどうかは未解決である.

\section{Other Topics}

5～7節での反自己双対Yang-Mills方程式の
解構成法はリダクションを通じて
そのまま低次元に応用される．
Yangの方程式(\ref{yang})を素朴に次元還元すると
Wardのカイラル模型・調和写像の方程式が得られる.
非自明なものとしては，例えば
Ernst方程式へのリダクションがある.
($z=x+\mathrm{i} t,w=re^{\mathrm{i}\theta}$として
$X=\partial_t, Y=\partial_\theta$
の方向に次元還元する.) 
これらに同様の解構成法が適用可能である. 
Ernst方程式は4次元Einstein方程式の
定常軸対称な真空解を与えるため，
5節のB\"acklund変換を適用して
ブラックホール解を生成することができる. 
4節(a)の3つの例については
Penrose-Ward変換が逆散乱変換そのものとなる．

反自己双対Yang-Mills方程式階層も定義することができ，
低次元の可積分階層に帰着する.
Drinfeld-Sokolov階層の方程式系を導くこともできる
(4節のKdV, Boussinesqへのリダクションはその一例である). 
そこに現れるさまざまな可積分系としての要素に
ツイスター理論の解釈が与えられる.

場の理論の枠組みからははずれるが，
1次元(常微分方程式)へのリダクションにおいても
美しい結果がある. 特に, 
4次元の共形変換群に3次元の可換な部分群が
5つあることに着目し, 
それらを生成する3つのベクトル場の方向に
次元還元すると，
5つの部分群に対応して, 
Painlev\'e方程式
P$_{\mbox{\scriptsize{I,II}}}$, 
P$_{\mbox{\scriptsize{III}}}$, 
P$_{\mbox{\scriptsize{IV}}}$, 
P$_{\mbox{\scriptsize{V}}}$, 
P$_{\mbox{\scriptsize{VI}}}$ 
が見事に導出される\cite{MaWo}. 

このようにさまざまな種類の可積分性の起源が
ツイスター理論からある程度統一的に理解される.
ただ，説明のつかない話題もあり，佐藤理論の中心的存在である
KP方程式はいまだにリダクションの例が
知られていない\footnote{ゲージ場の要素に
微分作用素を許すようなリダクションからは
導かれるが，ツイスター対応がなく
Ward予想の一例とはみなされない.}.

\section{Perspectives}

このストーリーのさらなる展開や今後の展望は，
高次元化，非可換化も含め
Ward氏の解説\cite{HSW}の最後にまとめられている.
ツイスター対応が適用できる範囲で
4次元時空を一般化する方向性もある\cite{Calderbank}. 

反自己双対Yang-Mills方程式の高次元化については
昔から議論がある．最近の興味深い話題として，
Cherkis氏が，8次元の「反自己双対な」Yang-Mills方程式を
3,5,7次元に単純な次元還元を施し，それらの方程式同士に
Nahm変換的な双対関係
を予想している\cite{Cherkis}.
ツイスター理論に収まれば，
(もしかしたらKP方程式も含むような)非自明なリダクションから
より壮大な枠組みが提供されるかもしれない.

非可換化については2000年頃から研究が活発である\cite{Hamanaka}.
(非可換な可積分系としては例えば従属変数が
四元数に値をとる設定をイメージすればよい.)
Ward予想の非可換版も主要な具体例は出揃っている\footnote{
4節の非線形Schr\"odinger方程式の例は
順序をあえて区別して記載してあり，実はこのまま
非可換版の例になっている. ゲージ場のトレース部分
(ゲージ群の$U(1)$部分)が生じていることに注意.
(Tr$ A_z=\overline{\psi}\psi-\psi\overline{\psi}\neq 0$.)}.
解構成においては5節で述べたQuasideterminantが大いに活躍する.
非可換な場合もQuasideterminant 
の具体的な表式を書き下すことができ，
5節で紹介した恒等式だけでなく
非可換Pl\"ucker恒等式など興味深い関係式が存在する\cite{GGRW}.
非可換対称関数とのつながりもある\cite{GKLLRT}.
これらは実際の計算に非常に有用であり, 
可換な場合より(普通の行列式を用いる場合より)
簡単に証明が完結する.
例えば，非可換な場合の
Atiyah-Ward仮設解はQuasideterminantを用いた表記
(\ref{AWsol})が答えである\footnote{ 
実は5節の議論は, Quasideterminantを用いた非可換での結果\cite{GHN}の
可換極限である.}. 
Yangの行列は, (\ref{AWsol})を (\ref{J})に代入すれば求まるが, 
5節で紹介した恒等式を用いると
以下のような非常に簡明で見通しの良い形にすぐにたどりつく\cite{GHN}：
\begin{eqnarray*}
&&J_l=
\begin{pmat}|{||..|}|
\fbox{0}&-1&0&\cdots&0&\fbox{0}\cr\-
1&\varDelta_0&\varDelta_{-1}&\cdots&\varDelta_{1-l}&\varDelta_{-l}\cr\-
0&\varDelta_1&\varDelta_0&\cdots&\varDelta_{2-l}&\varDelta_{1-l}\cr
\vdots&\vdots&\vdots&\ddots&\vdots&\vdots\cr
0&\varDelta_{l-1}&\varDelta_{l-2}&\cdots&\varDelta_{0}&\varDelta_{-1}
\cr\-
\fbox{0}&\varDelta_{l}&\varDelta_{l-1}&\cdots&\varDelta_1&\fbox{$\varDelta_{0}$}\cr
\end{pmat}.
\end{eqnarray*}
ボックスが4個あるように見えるが，
４隅の要素を$2\times 2$行列として
一つのボックスで囲っているつもりである.
(要素として非可換なものも許されるので行列が入っても構わない.)
Quasideterminantはもともと非可換行列式の統一的理解のため
に定義されたものであるが，非可換ソリトンの記述に
ぴったりはまっているように思える. 
低次元可積分系でも活躍することが分かっており\cite{EGR,GiNi}
可積分系のより本質的な新しい定式化が
予感される. 

対称性という観点に戻ってみると，
反自己双対Yang-Mills方程式の時空対称性やゲージ対称性は, 
低次元可積分系へのリダクションの分類指標を与えているが
可積分性の直接の起源ではない．
時空対称性が共形対称性まで大きくなったことで
ツイスター理論の取扱いが可能になったのが一つの理由である．
リダクションの過程でツイスター対応が
つぶれないことも条件である. 
この意味で高い時空の対称性が, ツイスター理論を経由して
可積分性と関わっている.  
対称性と可積分系の関係はもっと直接見えないだろうか？

不定値計量の反自己双対Yang-Mills方程式は
実は (世界面の超対称性が通常の2倍ある) N=2弦理論\cite{Ooguri}
と深く関わっている\footnote{N=2弦理論の標的空間は
不定値計量の4次元空間であり，
その上の弦の場の理論の運動方程式(BPS方程式ではない)が
反自己双対Yang-Mills方程式などを与える.}. 
理解の鍵は可積分系・ツイスター理論の
弦理論的解釈にあるのかもしれない.



\end{CJK}

\end{document}